\documentclass[conference]{IEEEtran}
\IEEEoverridecommandlockouts
\usepackage{cite}
\usepackage{amsmath,amssymb,amsfonts}
\usepackage{algorithmic}
\usepackage{graphicx}
\usepackage{textcomp}
\usepackage{xcolor}
\usepackage{makecell}
\usepackage{multicol}
\usepackage{multirow}
\usepackage{verbatim}
\def\BibTeX{{\rm B\kern-.05em{\sc i\kern-.025em b}\kern-.08em
    T\kern-.1667em\lower.7ex\hbox{E}\kern-.125emX}}

\begin{document}

\title{Few-shot Detection of Anomalies in Industrial Cyber-Physical System via Prototypical Network and Contrastive Learning\\
}

\author{
\IEEEauthorblockN{Haili Sun}
\IEEEauthorblockA{\textit{School of Cyber Science and Engineering} \\
\textit{Huazhong University of Science and Technology}\\
Wuhan, China \\
hailisun@hust.edu.cn}
\\
\IEEEauthorblockN{Lansheng Han$^{*}$ \thanks{* Corresponding Author}}
\IEEEauthorblockA{\textit{School of Cyber Science and Engineering} \\
\textit{Huazhong University of Science and Technology}\\
Wuhan, China \\
hanlansheng@hust.edu.cn}

\and

\IEEEauthorblockN{Yan Huang}
\IEEEauthorblockA{\textit{School of Artificial Intelligence and Automation} \\
\textit{Huazhong University of Science and Technology}\\
Wuhan, China \\
platanus@hust.edu.cn}
\\
\IEEEauthorblockN{Chunjie Zhou}
\IEEEauthorblockA{\textit{School of Artificial Intelligence and Automation} \\
\textit{Huazhong University of Science and Technology}\\
Wuhan, China \\
cjiezhou@hust.edu.cn}
}

\maketitle

\begin{abstract}
The rapid development of Industry 4.0 has amplified the scope and destructiveness of industrial Cyber-Physical System (CPS) by network attacks. Anomaly detection techniques are employed to identify these attacks and guarantee the normal operation of industrial CPS. However, it is still a challenging problem to cope with scenarios with few labeled samples. In this paper, we propose a few-shot anomaly detection model (FSL-PN) based on prototypical network and contrastive learning for identifying anomalies with limited labeled data from industrial CPS. Specifically, we design a contrastive loss to assist the training process of the feature extractor and learn more fine-grained features to improve the discriminative performance. Subsequently, to tackle the overfitting issue during classifying, we construct a robust cost function with a specific regularizer to enhance the generalization capability. Experimental results based on two public imbalanced datasets with few-shot settings show that the FSL-PN model can significantly improve F1 score and reduce false alarm rate (FAR) for identifying anomalous signals to guarantee the security of industrial CPS.
\end{abstract}


\section{Introduction}
Cyber-Physical Systems (CPS) are large, distributed, heterogeneous and multidimensional intelligent systems, integrating communication, computation and control to realize close combination and coordination of physical and software resources \cite{b1}. These systems offer rich functionalities that are widely applied in industries such as healthcare, critical infrastructure and intelligent transportation, which are vulnerable to cyber-physical attacks due to their complexity and heterogeneity \cite{b2}. If they are compromised, it may poses disastrous economical and environmental consequences. For example, in 2000, the attack on the SCADA system at the Maroochy wastewater treatment plant in Queensland, Australia, resulted in 750,000 gallons of effluent flowing out, causing loss of marine life and endangering public safety \cite{b3}. In 2015, BlackEnergy malware attack on Ukraine's electricity infrastructure caused power outages in more than half of the country \cite{b4}. Stuxnet \cite{stuxnet5} destroyed roughly a fifth of Iran’s nuclear centrifuges by causing them to spin out of control. Therefore, ensuring the safety and security of CPS is of paramount importance.

Anomaly detection is an important technique for dealing with cyber-physical attacks. It aims to distinguish normal and abnormal system behaviors from multivariate system data such as network traffics, sensor values and system logs generated by CPS. For anomaly detection, algorithms based on deep learning (DL) have been extensively explored. For instance, Zhou et al. \cite{b3} proposed an intelligent anomaly detection model VLSTM using encoder-decoder neural network with a variational reparameterization schema for intelligent industrial application. Kasongo et al. \cite{b6} designed a classifier based deep long-short term memory for detecting intrusions in wireless networks. However, these anomalies are usually sparse in real world which results in the lack of sufficient labeled samples for model training. This cause great challenge for most of the existing approaches which mainly rely on rich labeled data when handling anomaly detection task.

Few-Shot Learning (FSL) can enable models to distinguish novel categories with limited number of labeled examples. It aims  to tackle issues on lacking adequate data samples with supervised information. In recent years, some works based on FSL have been proposed for industrial anomaly detection. Zhou et al. \cite{b7} constructed a Siamese neural network for detecting industrial anomalies to solve the overfitting issue. Huang et al. \cite{b8} proposed a gated network structure to detect new anomaly types by aggregating seen anomaly types and unseen types in few-shot settings. However, their detection performance is not satisfactory due to they have no specialized feature module for extracting more distinctive features and complex relations from industrial CPS data. Moreover, recent studies have also shown that a well-designed feature extractor is more powerful than model using only a complex meta-learning algorithm for few-shot classification tasks \cite{b9}. Additionally, none of them noticed that both the extractor and classifier may overfit in limited sample settings.

Therefore, to address the above challenges, we present a few-shot learning method, named FSL-PN, for anomaly detection in industrial CPS with few labeled data. In particular, to extract meaningful features from high-dimensional samples, we design a lightweight efficient feature extractor consisting of four residual blocks. To further enhance the discriminative ability, we introduce a contrastive loss to guide the training process of the extractor for obtaining a tight intra-class and sparse inter-class feature space. Finally, we construct a classifier based on prototypical network for abnormality identification task. The abnormality detection is achieved by finding the nearest prototype in the feature space. Moreover, to alleviate the overfitting problem and improve its generalization ability of the classifier in the few sample scenario, we also add a regularizer based on the distance between sample features and prototype into the classifier framework.

The main contributions of this paper are as follows:
\begin{itemize}
\item We propose a two stage few-shot learning anomaly detection model based on prototypical network, named FSL-PN, is constructed based on prototypical network and constrastive learning to deal with the limitation of labeled samples in industrial CPS. Furthermore, to improve the generalization ability of the model, a regularizer is designed based on the distance from the features to the prototype, so that similar samples are distributed more compactly in the feature space.
\item The temperature coefficient is dynamically adjusted according to label information via supervised contrastive learning, thus to generate embedding space which are compact within clusters and sparse between clusters. In addition, We also propose a robust cost function SPInfoMax for detecting anomalies, which are identified by maximizing mutual information between samples and prototypes. Furthermore, to improve the generalization ability, a regularizer is designed based on the distance from the features to the prototype, so that similar samples are distributed more compactly in the feature space.
\item To improve the detection rate, we construct a lightweight feature extractor to encode meaningful features from high-dimensional data. Moreover, we also design a contrastive loss for the extractor to learn more discriminative features from original data which can further promote the detection robustness.
\item Extensive experiments are conducted on two public datasets including UNSW\underline{ }NB15 and NSL\underline{ }KDD. Comparing with existing optimal few-shot learning method (FSL-SCNN) and other methods including one-shot support vector machine (OS-SVM), random forest, naive bayes, and VLSTM, the proposed method is superior to existing methods.
\end{itemize}

\section{Related Work}
In recent years, researchers have focused on addressing the vulnerability and security of CPS, which has been implemented in various applications in industry, ranging from data acquisition, monitoring, and industrial control systems. There exist many kinds of cyber-physical attacks that can compromise the security and reliability of CPS. For example, focusing on the false-data injection attacks, Beg et al. \cite{b10} designed a detection architecture to identify changes from inferred candidate invariants, which are inferred by Simulink/Stateflow diagrams in CPS. Sun et al. \cite{b11} foucsed on mitigating the impact of Dos attacks, and proposed a resilient model predictive control strategy with a dual-mode algorithm to ensure exponential stability of CPS. To detection zero dynamics and covert attacks, Heohn et al. \cite{b12} Designed a modulation matrix and inserted it in the path of the control variables for revealing these two attacks in CPS.

Particularly, anomaly detection algorithms have been extensively analyzed and studied for the security and reliability of industrial CPS. It is crucial to develop appropriate detection architectures to identify attacks on these systems under different network scenarios. Several artificial intelligence based methods have been designed to protect CPS against cyber-physical attacks, such as error diagnosis, attack detection, and fault-tolerant control \cite{b13}. To identify cyber attacks in industrial control systems, reference \cite{b14} used deep learning algorithms to model the statistical deviation of observed and predicted values. 1D CNN networks was utilized to detect attacks in SWAT dataset. Pearce et al. \cite{b15} proposed a bidirectional runtime enforcement to mitigate the damages posed by compromised controllers in CPS.  The reference \cite{b16} proposed a dual DL monitoring system using energy auditing data to identifying cyber-physical attacks in the IoT environment. They developed an aggregation-disaggregation structure to model the system behaviors.  To identify anomalies from raw data of intelligent industrial application, Zhou et al. \cite{b3} proposed a variational long short-term memory model based on reconstructed feature representation. Obviously, the aforementioned approaches shows that DL-based methods have been successfully used to identify cyber-physical attacks in industrial CPS. However, as traditional supervised learning methods typically rely on rich labeled data and prior knowledge, they may have difficulty in effectively detecting novel anomalies from few labeled samples in smart industrial environment.

\textbf{Contrast learning for anomaly detection:} Contrast learning is widely used in representation learning in computer vision \cite{b24,b25,b26,b27}. In view of its powerful representational learning ability, many people transfer it to the field of anomaly detection \cite{b28,b29,b30}. Kopuklu et al.\cite{b28} Adopt supervised comparative learning. for detecting anomalous driving. In the discriminant phase, they calculate the mean of all normal samples as the center and identify anomalies by comparing the distance of the test sample to that center with the threshold value. However, he can only identify one classification problem, and the quality of the detection results depends on the suitability of the threshold. On the basis of traditional contrast learning, the anomaly detection of mask contrast learning is proposed by designing a task-specific variant for CV field \cite{b30}. A Class condition mask was designed to dynamically adjust the temperature coefficient of contrast loss according to the pseudo-label of samples. However, their method involves transformation enhancement of the image, such as grayscale processing, rotation, etc., which is not suitable for processing industrial CPS data. for out-of-class detection task with tabular data, Shenkar et al.\cite{b29} proposed to maximize the mapping relationship of mutual information between each sample and its shielded part through comparative loss learning. After that, the learned mapping relationship is used to calculate the anomaly score of the test sample for anomaly judgment.

\textbf{Few-shot learning for anomaly detection:} Few-shot learning is a new transfer learning paradigm \cite{b17}. By reusing the transferable knowledge of existing classes, it can identify novel categories from a limited number of labeled samples \cite{b18}. Chowdhury et al. \cite{b19} proposed a deep CNN to extract feature representations, which were fed to the integration of a support vector machine and one-nearest neighbor classifier for identifying intrusions. To address the flare soot density few shot problem, Gu et al. \cite{b20} designed a recognition network for industrial safety and environment protection. To handle the imbalanced data problem, Huang et al. \cite{b8} constructed a few-shot learning model with a gated network structure to detect new anomalies by aggregating seen anomaly types and unseen types. To enhance the security of CPS, Zhou et al. \cite{b7} introduced a few-shot learning model based on Siamese CNN named FSL-SCNN for intelligent anomaly detection in industrial CPS.

Prototypical network (PN) [22] is a popular few-shot learning method, given its good learning ability, it has to face a problem: the prototype calculated based on the mean of a small number of samples is inaccurate. As mentioned above, although previous researches could handle the few-shot detection issues, their performance may be limited by the lack of a specifically designed feature extract module and the model may suffer from overfitting problem in both the feature extraction and classification stages in few-shot learning scenarios. Therefore, different from existing approaches, to promote the detection accuracy, we design a dedicated feature extractor and introduce the contrastive loss to extract more discriminative high-dimensional features from original input and design a robust prediction loss. Furthermore, we built a few-shot anomaly detector based on PN. In order to make the prototypes more close to the real sample center, we propose to maximize the mutual information between the samples and the prototypes, so as to narrow the distance between the samples and the corresponding prototypes and enlarge the distance between the samples and other prototypes. As a result, we can obtain a prototype-centric class which is more compact within the class and with a larger margin between classes in the embedding space. In addition, to deal with the overfitting problem of the classifier, we introduce a regularizer into the prediction loss to improve the generalization ability of the model.

\section{Few-Shot Learning Methodology}
\subsection{Problem Definition and Formalization}
In the industrial CPS anomaly detection scenario, given a data set \textit{D} containing both normal and abnormal samples. To describe the few-shot learning scenario, we assume that abnormal data scale is far less than normal data one. In order to form the \textit{C}-way \textit{N}-shot learning task, we randomly select  categories from \textit{D}, each category contains \textit{N} samples to form the support set $Tr_{S}$ and the corresponding unseen query set $Tr_{Q}$, which indicates other samples of the same categories \textit{C}  in each training episode.

\subsection{Proposed Framework for Anomaly Detection}
The proposed model FSL-PN aims to address the issue on lacking of sufficient abnormal data in our detection task. Unlike traditional classification models in industry, we develop a feature extractor and introduce the contrastive loss for extracting more discriminative latent representations from raw input samples.

\begin{figure}[htbp]
\centerline{\includegraphics{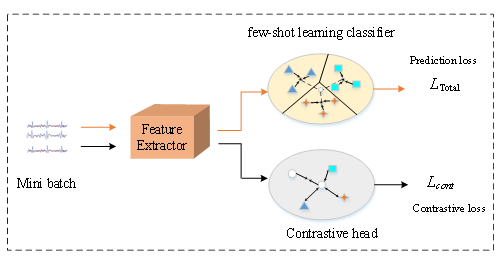}}
\caption{The proposed anomaly detection framework.}
\label{fig1}
\end{figure}

Then, we construct a classifier based on prototypical network to deal with the FSL problem, hence the novel categories can be recognized even with only a few labeled data. The overall framework of FSL-PN for anomaly detection in CPS is shown in  Fig. \ref{fig1}.

As shown in Fig. \ref{fig1}, the FSL-PN has three main components: an feature extractor, a contrast head and a classifier. The first
one is designed to extract meaningful features from input samples, the second
one is used to guide the training process of the extractor so that it will
acquire powerful feature extraction capability, and the last classifier tries to distinguish whether an input sample is abnormal or not.

The feature extractor $f$ is constructed based on residual blocks, which extends the neural network to a very deep structure by adding shortcut connections in each residual block to allow the gradient flow directly through the underlying layer. It stacks four residual blocks, each block contains two convolution layers. To alleviate the overfitting issue, we also exclude any pooling operation like ResNet. After the residual blocks, a global average pooling layer is used to pool the features instead of a fully connected layer for reducing the number of weights. Subsequently, these features are fed into the contrastive head or classifier. The specific structure of the extractor is shown in the left part of Fig. \ref{fig1}. For simplicity, we omit batch-normalization and activation layers. 

The main purpose of the contrastive head module is to guide the extractor training process, which is instantiated as a multi-layer perceptron with a single hidden layer and an output layer of size 128. Subsequently, the output vector is normalized to lie on the unit hypersphere, thus the distances between two vectors can be measured by their inner product. Given the extracted feature representation \textit{f(x)} of an input sample \textit{x}, it will be mapped as \textit{z} = \textit{MLP}(\textit{f}(\textit{x})) by the contrastive head. Then, the distance between two input samples \textit{$x_i$} and \textit{$x_j$} can be formalized as:

\begin{equation}
D(z_{i},z_{j})= z_{i} \cdot  z_{j} \label{eq}
\end{equation}
\begin{equation}
z_{i}= MLP(f(x_{i})),z_{j}= MLP(f(x_{j}))
\end{equation}
\subsection{Robust Cost Function Design}
The trainable parameters of proposed framework consist of two parts: one is the parameter of the feature extractor, denotes as $\theta$ , and the other is the parameter of the classifier, denotes as $\varphi$. Furthermore, to extract discriminative features from the original input, supervised contrast loss is introduced to guide the training process of the feature extractor, which greatly facilitates the classification performance. In addition, to further reduce overfitting and false alarm rate (FAR), a regularizer is added in the design of the prediction loss function to improve the generalization ability of the model.

As mentioned above, the model is trained in two stages: supervised contrastive learning stage and few-shot classification stage. The first stage (shown by the black arrows in Fig. \ref{fig1}) aims to learn the encoder \textit{f} : \textit{x} $\rightarrow$ \textit{h}, which is the key component that maps the input samples into a low dimensional compact latent feature space while the second stage (shown by the orange arrows in Fig. \ref{fig1}) utilizes explicit few labeled information to fine-tune the classifier with the previously trained extractor. Noteworthyly, the extractor is frozen and taken as a fixed module during this stage. We believe that this is beneficial to avoid the overfitting issue of limited label data when identifying anomalies.

\textbf{Supervised Contrastive Learning Stage:}
This stage is a supervised instance-level classification task, which is designed to recognize latent fine-grained structure in the low dimensional feature space by separating the representations of different latent classes and aggregating those of the same latent classes simultaneously. Concretely, for a mini-batch of  \textit{N} features $\{{z_{i}, y_{i}}\}_{i= 1}^{N}$, where    $z_{i}= MLP(f_\theta(x_{i}))$ is a vector lied in a hypersphere mapped by the contrastive head as aforemetioned, and $y_{i}$ is the label of the ground truth, then the loss can be defined as:
\begin{equation}
    L_{cont}(\theta )= \frac{1}{N}\sum_{i= 1}^{N}L_{z_{i}}
\end{equation}
\begin{equation}
    L_{z_{i}}= \frac{-1}{\left| N_{y_{i}}\right|-1}\sum_{k\in N_{y_{i}}}^{}log\frac{exp(D(z_{i}, z_{k})/\tau )}{\sum_{q= 1}^{N}exp(D(z_{i},z_{q})/\tau)}\label{eq4}
\end{equation}

Here, $N_{y_{i}}$ is the set of indices of all samples with the same label as $y_{i}$ in an episode, $\left| N_{y_{i}}\right|$ is its cardinality, and  $\tau$ is a scalar temperature hyper-parameter.

As in \eqref{eq4}, $D(z_{i}, z_{k})$ denotes the distance between the $i-th$ and $k-th$ sample in the latent space and thus the objective of the above loss function is to reduce the
\begin{figure*}[htbp]
\centerline{\includegraphics[width=8in]{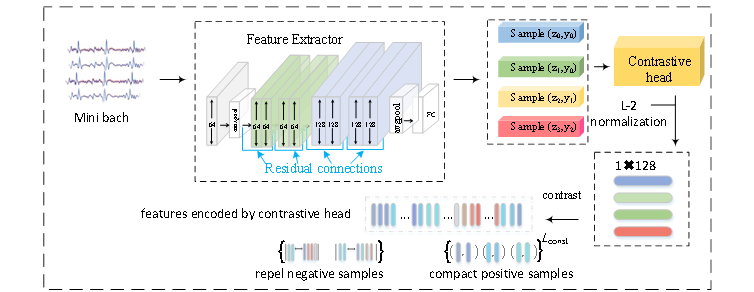}}
\caption{Supervised contrastive learning architecture.}
\label{fig2}
\end{figure*}instance-level distance between samples with the same label and enlarge those with different labels. Consequently, the samples of each category will form a tighter cluster and the boundaries between different clusters will be enlarged. This will facilitate the following classification task to calculate more accurate cluster center, i.e. class center. The training process of this task is shown in Fig. \ref{fig2}.

\textbf{Modification of temperature coefficient:}
The temperature coefficient is used to regulate the degree of attention paid to hard samples. The larger the temperature coefficient is, the more attention will be paid to other similar samples. But for supervised contrast learning, the category information is known. In order to learn the embedded space with clearer boundaries, that is, to make the model exclude non-homogeneous samples as much as possible, we inject tag-specific features into the existing contrast learning, so that it does not pay attention to the distinction between similar samples, but extracts the category-related features. The key component of the first phase of the model is class-information injection (CII), a simple and effective injection technique that adaptively determines rejection rates based on labels. The CII value can be defined as:

\begin{equation}
    CII=\left\{\begin{aligned}
 \beta & , & y_{i} =  y_{k},\\
 \tau & , & y_{i}\neq y_{k}. 
\end{aligned}
\right.
\end{equation}

where $0<\tau<\beta$. CII sets a larger temperature coefficient $\beta$ for the same kind of samples, so the strength of the anchor sample to repel the same kind of samples is smaller than that of the samples with different labels. Then the generated CII is divided by the similarity score, so the improved comparison loss is formulated as Eq.(6):

\begin{equation}
    \begin{aligned}
    L_{z_{i}} = \frac{-1}{\left| N_{y_{i}}\right|-1}\sum_{k\in N_{y_{i}}}^{}log\frac{exp(D(z_{i}, z_{k})/\tau )}{\sum_{q= 1}^{N}exp(D(z_{i},z_{q})/CII(i,q))}\label{eq5}
    \end{aligned}
\end{equation}

By setting a small penalty rate of $1/\beta$ for the same type of sample pairs, the unique representations of different samples in the same type of cluster can be learned, and can avoid the characteristics of the samples in this class to be too similar.

\textbf{Robust Cost Function based InforMax for Classification (CFD):}

We propose to maximize the mutual information between prototypes and samples. In this way, the sample of the query set is also used for prototype generation. Theoretically, incorporating query sets into fine-tuning prototypes is conducive to improving the accuracy of prototypes, because the more samples involved in prototype calculation, the more accurate the prototypes will be.

In view of the inaccuracy of prototypes based on the calculation of a small number of samples in prototype few-shot learning, the mutual information maximization (InfoMax) training objective is utilized to alleviate this problem via taking the normal sample distribution and abnormal sample distribution as a comparative view. The idea is simple: we want to maximize the mutual information between the sample and the owning prototype, while penalizing the abnormal sample.

Here, the task of anomaly detection is also formalized into a comparative learning framework through infoMax. The model is forced to learn the binary exclusion function L, which divides the embedded space according to the normal and abnormal data sets.

This approach can be simply extended to scenarios with multiple classifications. The introduction of a binary exclusion function for small sample detection in prototype network can not only alleviate the problem of prototype inaccuracy based on mean calculation, but also simplify the whole training mechanism. That is, $L$ can be approximated as the cross entropy realization of a simple InfoMax target. Thus, in the current anomaly detection problem, losses can be formalized as approximate mutual information between the query set sample and the prototype vector of the support set $I(C_i, Q)$ :

\begin{equation}
    \begin{aligned}
        I(C_i,Q) >= & E_n[log S(C_i, x_{normal})]+E_a[log(1-S(C_i, \\ 
        & x_{abnormal}))] + E_{abnormal}
    \end{aligned}
\end{equation}

where $E$ is the expectation. $x_{normal}$ and $x_{abnormal}$ are normal and abnormal samples in the query set, respectively. $x_n$ belongs to $Q_n$, $x_a$ belongs to $Q_a$, and ${Q_n,Q_a}$ belongs to $Q$. $S$ is a similarity rating layer.
Therefore, InfoMax loss of samples and prototypes can be defined as binary cross entropy loss formulated as below:
\begin{equation}
    \begin{aligned}
        L(C_{i},Q,\varphi) = &\frac{1}{|Q_{n}|}\sum_{x_{n}\in Q_{n}}^{}log S(C_{i}, x_{n}) + \\
        & \frac{1}{|Q_{a}|}\sum_{x_{a}\in Q_{a}}^{}log(1-S(C_{i}, x_{a}))
    \end{aligned}
\end{equation}

where $\varphi$ is the parameter of the classifier. 
Intuitively speaking, maximizing mutual information is equivalent to narrowing the samples and the belonging prototypes, and amplifying the distance between the sample and other prototypes, which can generate a more compact embedding space within a class and a larger margin between various types. Thus reducing the deviation of the prototype from the real center of the sample. 

\textbf{Regularization Term:} Getting more accurate prototypes for better classification accuracy by minimizing the formula $L(C, Q)$.

However, in the few-shot settings, directly minimizing this prediction loss may lead to overfitting. To overcome this issue, we introduce the distance between samples and corresponding prototype as a regularizer into the prediction loss to improve the generalization ability of the model. This distance-based regularizer is defined as:
\begin{equation}
    L_{regu}(\varphi)= d(F_{\varphi}(f(x),C_i)
\end{equation}

Finally, the overall prediction function is: 
\begin{equation}
    L_{CFD}=L(C,Q, \varphi) + \alpha L_{regu}(\varphi)
\end{equation}
where $\alpha$ is a balance coefficient to control the weights of the regularizer. It also can be seen as maximum likelihood regularization \cite{b21}, extensively used in pattern recognition.

$L_{regu}$ can further improve the performance of the classifier because: (1) it pulls the features of the samples near their corresponding prototypes, so that the sample features of the same class are distributed more compactly, which can implicitly increase the distance between classes and therefore boosting the classification; (2) the prediction loss emphasizes the separability of the representation and the regularizer emphasizes the tightness of the representation, thus by combining them together, we can learn intra-class tightness and inter-class separable representations, which are more robust and better suited for close and open set problems.

\begin{figure}[htbp]
\centerline{\includegraphics[width=3.8in]{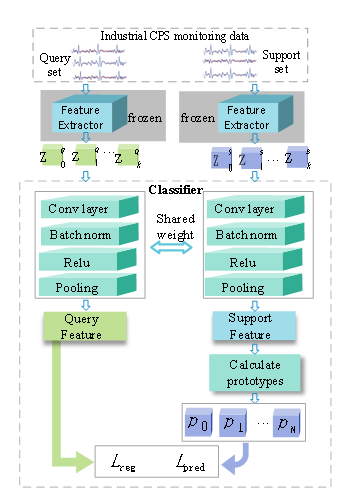}}
\caption{Few-shot classification architecture.}
\label{fig3}
\end{figure}

\textbf{Few-shot Classification Stage:}
As shown in Fig. \ref{fig3}, 
the goal of this stage is to train the classifier using the features extracted by the trained extractor. In this stage, to avoid overfitting, the parameters of the extractor are fixed and only the ones of the classifier are trained.

\textbf{Prototype Generation:} The sampled data in both $Tr_{S}$ and $Tr_{Q}$ of an episode $T$ are used. The extractor first maps all samples into the latent feature space. Then, all feature representations of the same class $M_{i}$ in support set $Tr_{S}$ are aggregated into a prototype $C_{i}$. Typically, it is completed by computing the average of all support features:
\begin{equation}
    C_{i}= \frac{1}{\left|Tr_{S}\right|}\sum_{(x,y)\in Tr_{S}\&[y=y_{i}]}^{}F_{\varphi}(f(x))
\end{equation}
where $F_{\varphi}$ denotes the convolution layers and $\varphi$ is the hyper-parameter of the classifier.

As shown in Fig.3, the goal of this stage is to train the classifier using the features extracted by the trained extractor. In this stage, to avoid overfitting, the parameters of the extractor are fixed and only the ones of the classifier are trained. 

\textbf{Inference:}
In our framework, the similarity of samples and prototypes can be measured by the distance between them. Thus, the probability $p(x\in y_{i}|x)$ that $x$ belongs to the prototype $C_{i}$ should be proportional to the negative of the distance between them. To satisfy the property that probability must be normalized and non-negative, we define $p(x\in y_{i}|x)$ as:
\begin{equation}
    p(x\in y_{i}|x)= \frac{exp(-d(F_{\varphi}(f(x)), C_{i}))}{\sum_{j}^{}exp(-d(F_{\varphi}(f(x)), C_{j}))}
\end{equation}
where $d(F_{\varphi}(f(x)), C_{i})= \left\| F_{\varphi}(f(x)) - C_{i}\right\|_{2}^{2}$ represents the Euclidean distance between sample $x$ and class $y_{i}$. Based on this probability, we can further define the optimization objective of the classifier as negative log-probability, that is:
\begin{equation}
    L(\varphi)= -log p(x\in y|x)\label{eq6}
\end{equation}

Equation \eqref{eq6} is used to measure the prediction accuracy. By minimizing this loss, the classifier can be trained to be able to correctly distinguish samples. 

\section{Experiment and analysis}

In this section, we conduct experiments to evaluate the effectiveness of our proposed method for anomaly detection in industrial CPS, comparing with a few-shot learning method and other similar mechanisms based on two public datasets.
\subsection{Dataset and Experimental Settings}
\textbf{Dataset:} to verify the effectiveness of our proposed model, we conducted a plenty of experiments on two public datasets, i.e. UNSW-NB 15 and NSL-KDD. The former was created by the Intelligent Security Group of Australia using the IXIA PerfectStorm Tool. It consists of network packets with 49 features, including synthetic contemporary attack behaviors and real modern normal activity packets. These attack behaviors contain nine types of network attacks including Backdoors, Generic, Analysis, Shellcode, Dos, Fuzzers, Worms, Reconnaissance and Exploits. The latter, i.e. NSL-KDD is an improved version of KDD CUP99 datasets, which aims to solve the inherent problems of the KDD CUP99 datasets through collecting data from DARPA '98 IDS Evaluation Program. The training and test set of NSL-KDD contain 125973 and 22544 records respectively, each with 41 characteristics and corresponding labels. Each set contains four simulated attacks. The details of the two datasets are shown in Table \ref{tabd}.
\begin{table}[htbp]
\caption{ Details of the two datasets}
\begin{center}
\resizebox{.95\columnwidth}{!}{
\begin{tabular}{ccccc}
\hline
\textbf{Dataset}&\makecell{\textbf{Train}\\\textbf{records}}&\makecell{\textbf{Test}\\\textbf{records}}&\textbf{Features}&\textbf{Attack behaviours} \\
\hline
UNSW\_NB15&175341&82332&49&\makecell{Backdoors, Generic,\\ Analysis, Shellcode, Dos,\\Fuzzers, Worms,\\Reconnaissance, Exploits}\\
\hline
NSL\_KDD&125973&22544&41&\makecell{User to Root Attack,\\Probing Attack,\\Denial of Service Attack,\\Remote to Local Attack}\\
\hline
\end{tabular}
}
\label{tabd}
\end{center}
\end{table}

\textbf{Feature Selection:} We conduct feature selection and normalization for the above UNSW\_NB15 and NSL\_KDD datasets.
Feature selection is necessary for reducing computation, avoiding overfitting and improving detection performance. 
To this end, a Python library, named Featurewiz, was used for selecting high quality features from original data set, as its excellent performance in main Data science competitions. 
The feature selection process has two steps: first, SULOV method (which is based on the MRMR algorithm) is used to select the variable with high mutual information score and minimum correlation. The results of this step are shown in Figure 4. Figure 4 shows how the SULOV algorithm defines which features should be retained based on their mutual information score (MIS) with the pre-defined target. Lines represent the correlation between two features; thicker lines represent higher correlation. Circles represent the MIS, The larger the circle, the higher the MIS, and therefore more important for the pre-defined target. Features are selected based on the higher MIS and lower correlation with others. Second, with the features obtained by above step, use Recursive XGBoost to create model and eliminate features preventing underfeeding and overfeeding.
Figure 5 shows its execution process. Finally, Featurewiz outputs the selected features ranked by their MIS to the pre-defined target.
After feature selection operation, the number selected features in the two datasets are thirteen and fifteen.

\begin{figure*}[htbp]
\centerline{\includegraphics[width=6in]{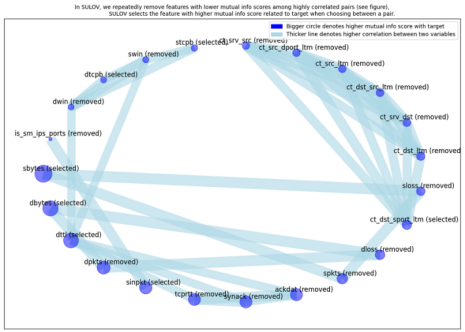}}
\caption{Graphic generated after the SULOV algorithm execution of Featurewiz on UNSW\_NB15 dataset.}
\label{fig-SULOV-unswnb15}
\end{figure*}

\begin{figure}[htbp]
\centerline{\includegraphics[width=3.8in]{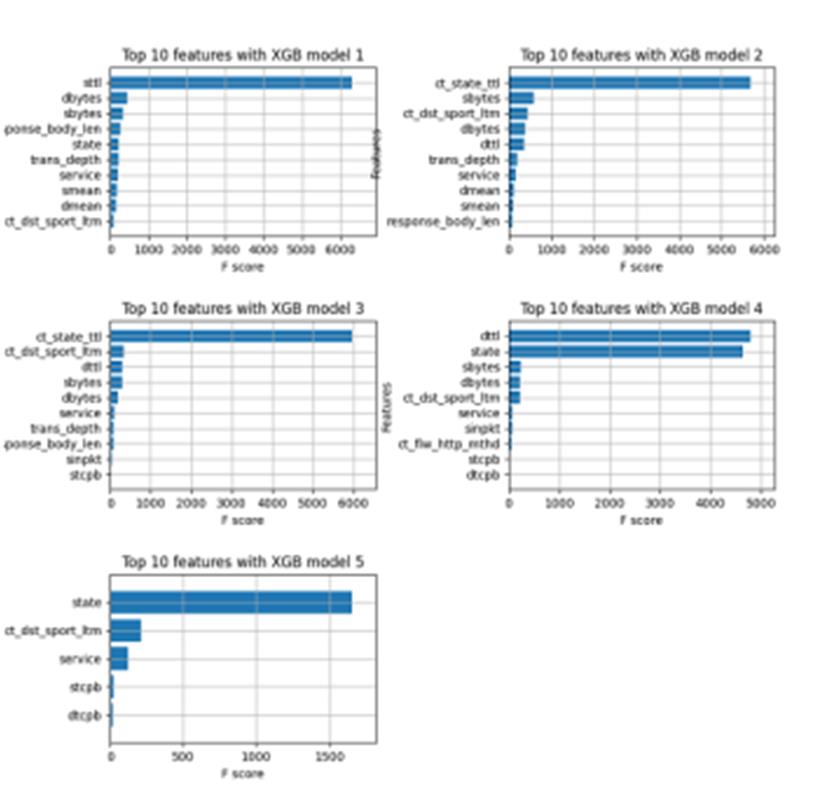}}
\caption{The execution process of the Recursive XGBoost step in Featurewiz on UNSW\_NB15 dataset.}
\label{fig-XGBoost-unswnb15}
\end{figure}

Since each metric of the network package has different magnitudes, the data values vary greatly, which will affect the detection results. Therefore, in order to eliminate the influence of the scale between different metrics, $L_2$ normalization was conducted on the data before training, with each sample (vector) scale as a vector of unit norm. Specifically, given a sample $x=[x_1,x_2,...,x_n]$, the normalized result will be $x/\sqrt{\sum_{i=1}^{n}x_i}$.

\textbf{Experimental Settings:} To validate the performance of our FSL-PN , a few-shot learning, a DL-based and several widely used traditional machine learning algorithms were selected as the baseline methods. Specifically, FSL-SCNN \cite{b7}, a Siamese CNN-based method; VLSTM \cite{b3}, a DL-based model, and three traditional machine learning algorithms such as OS-SVM, random forest (RF), and Naïve Bayes (NB) are compared in this paper. In addition, a non-machine learning time series analysis algorithm, named TSA, is involved for comparison evaluation as well.

We apply and calculate four widely used metrics, precision, recall, F1 and FAR depending on whether normal and abnormal samples have been recognized correctly or not, to prove the detection performances of the above methods.

\subsection{Anomaly detection performance evaluation}

The classifier of the model is implemented based on the Prototypical Network (PN) structure \cite{b22} . We set the initial learning rate to 0.001, choose stochastic gradient descent (SGD) as the optimizer, and iterate 1000 to train the model. The proposed model did the 2-way $k$-shot (=10, 5, 3, 2) detection tasks. The experimental results are shown in Table \ref{tab2}. 
\begin{table}[htbp]

\caption{ Detection performance of FSL-PN on UNSW\_NB15 and NSL\_KDD datasets}
\begin{center}
\resizebox{.95\columnwidth}{!}{
\begin{tabular}{cccccc}
\hline
\textbf{Datasets}&\makecell{\textbf{2-way}\\\textbf{$k$-shot}}&\textbf{precision}&\textbf{recall}&\textbf{F1}&\textbf{accuracy} \\
\hline
\multirow{4}{*}{UNSW\_NB15}&10-shot&93.61&94.5&93.74&93.65 \\
&5-shot&93.16&95.6&93.76&93.6 \\
&3-shot&94.57&95.67&94.17&94 \\
&2-shot&\textbf{95.17}&\textbf{97.5}&\textbf{95.57}&\textbf{95.25} \\
\hline
\multirow{4}{*}{NSL\_KDD}&10-shot&97.69&95.6&\textbf{96.36}&\textbf{96.65} \\
&5-shot&96.37&\textbf{96.4}&96.15&96.1\\
&3-shot&\textbf{97.92}&95&95.62&96.17\\
&2-shot&97.83&96&96.27&96.5\\
\hline
\end{tabular}
}
\label{tab2}
\end{center}
\end{table}
According to Table \ref{tab2}, it is observed that the detection performance of the model is not decreased but significantly improved as the training sample decreases on UNSW\_NB15 dataset. Especially for the 2-shot task, the recall metric of the model reached 97.5\%, which is 3\% higher than that of the 10-shot task. This indicates that the proposed model is indeed suitable for handling few labeled sample anomaly detection task. In addition, for the NSL\_KDD dataset, the performance of the model changes slightly as the number of samples changes. This may be due to the fact that there still exist some problems discussed by McHugh et al. \cite{b23}.

To further demonstrate the detection performance of the FSL-PN, comparative experiment with earlier selected benchmark models are conducted on UNSW\_NB15 dataset and the results are shown in Table \ref{tab3}. It is observed that the FSL-PN has achieved the best results in precision, F1 score and FAR at 97.5\%, 95.97\% and 3.5\%, respectively, promoting 6.9\%, 2.37\% and 1.2\% improvement than the subprime model. This indicates that through the well-developed extractor with the contrastive loss and the robust cost function  designed in our model, the FSL-PN can not only identify outliers from normal patterns efficiently but also reduce the FAR in the few-shot learning scenario. In particular, the reduction in FAR is of great significance for industrial applications.

\begin{table}[htbp]
\caption{ comparisons of detection performance on unsw\_nb15 dataset (\%)}
\begin{center}
\resizebox{.95\columnwidth}{!}{
\begin{tabular}{cccccc}
\hline
\textbf{Methods}&&\textbf{precision}&\textbf{recall}&\textbf{F1}&\textbf{FAR} \\
\hline
\multicolumn{2}{c}{TSA$^{\mathrm{a}}$}&87.10&87.00&87.00&15.7 \\
\multicolumn{2}{c}{OS-SVM}&88.60&	91.80&	90.20&	8.5\\
\multicolumn{2}{c}{RF}&	73.30&	97.40&	83.70&	18.1\\
\multicolumn{2}{c}{NB}&	70.60&	74.60&	72.50&	25.8\\
\multicolumn{2}{c}{VLSTM$^{\mathrm{b}}$}&	86.00&	97.80&	90.70&	11.7\\
\multicolumn{2}{c}{FSL-SCNN$^{\mathrm{a}}$}&	90.60&	96.80&	93.60&	4.7\\
\hline
\multirow{4}{*}{FSL-PN}&$\alpha=0$&	94.83&	96.00&	94.20&	8.0\\
&$\alpha=0.1$&	97.33&	96.00&	95.83&	4.0\\
&$\alpha=0.01$&	96.83&	95.00&	94.90&	4.5\\
&$\alpha=0.001$&	97.50&	96.00&	95.97&	3.5\\
\hline
\multicolumn{6}{l}{$^{\mathrm{a}}$The results come from \cite{b7}.} \\
\multicolumn{6}{l}{$^{\mathrm{b}}$The results come from \cite{b3}.}
\end{tabular}
}
\label{tab3}
\end{center}
\end{table}

\begin{figure}[htbp]
\centerline{\includegraphics[width=0.4\textwidth,height=0.4\textwidth]{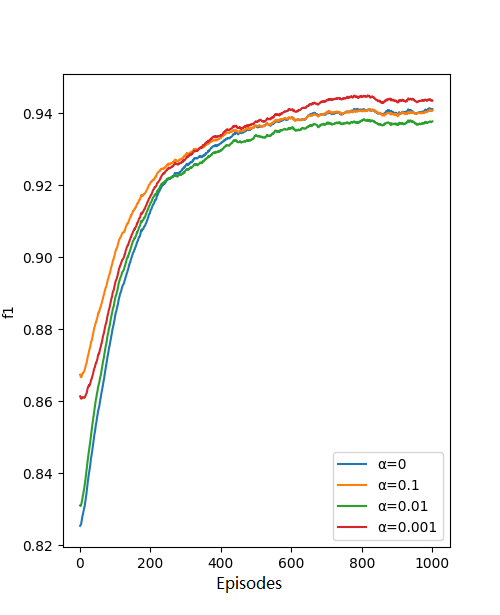}}
\caption{The F1 scores with increasing number of episodes for our proposed approach using regularizer with different coefficient $\alpha$.}
\label{figa_f1}
\end{figure}

\begin{figure}[htbp]
\centerline{\includegraphics[width=0.4\textwidth,height=0.4\textwidth]{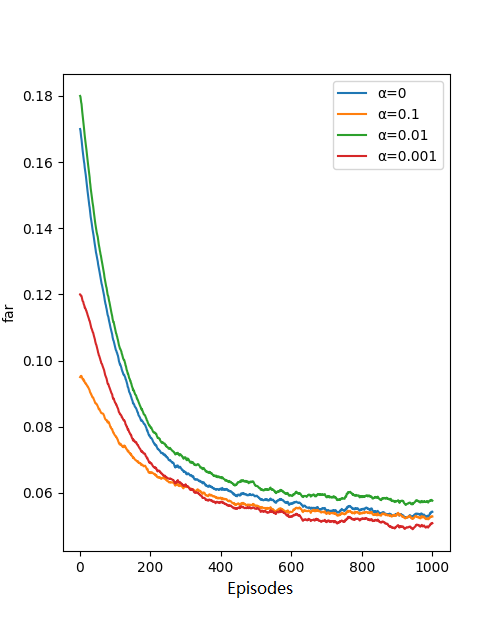}}
\caption{The False alerm rate (far) with increasing number of episodes for our proposed approach using regularizer with with different coefficient.}
\label{figa_far}
\end{figure}

\begin{table}[htbp]
\caption{2-way 2-shot ablation study on UNSW\_NB15 dataset (\%)}
\begin{center}
\begin{tabular}{ccccc}
\hline
\textbf{Method}&\textbf{pre}&\textbf{rec}&\textbf{F1}&\textbf{acc} \\
\hline
PN &	93&	92&	90.67&	90.75\\
F($\cdot$) + linerclassifier&	94.37&	93.12&	93.01&	93.12\\
F($\cdot$) + PN&	94.83&	96.9&	95.85&	94.25\\
F($\cdot$) + PN + CII&	95.25&	97.2&	96.22&	94.75\\
F($\cdot$) + PN +CII + SPinfomax (ours)&	\textbf{95.88}&	\textbf{97.6}&	\textbf{96.73}&	\textbf{95.25}\\
\hline
\end{tabular}
\label{tab4}
\end{center}
\end{table}

\begin{table}[htbp]
\caption{2-way 2-shot ablation study on NSL\_KDD dataset (\%)}
\begin{center}
\begin{tabular}{ccccc}
\hline
\textbf{Method}&\textbf{pre}&\textbf{rec}&\textbf{F1}&\textbf{acc} \\
\hline
PN	& 	94.5&	93&	93.2&	94.25\\
F($\cdot$) + linerclassifier&	87.77&	83.5&	82.43&	83.5\\
F($\cdot$) + PN& 97.83&	96&	96.27&	96.5\\
F($\cdot$) + PN + CII&	98.25&	96.87&	97.68&	96.80\\
F($\cdot$) + PN +CII + SPinfomax (ours)&	\textbf{98.68}&	\textbf{97.40}&	\textbf{98.04}&	\textbf{97.25}\\

\hline
\end{tabular}
\label{tab5}
\end{center}
\end{table}

\begin{table}[htbp]
\caption{Influence of Resnet's convolution layer number on detection result (\%)}
\begin{center}
\begin{tabular}{cccccc}
\hline
\textbf{Method}&\textbf{Conv Layer Num}&\textbf{pre}&\textbf{rec}&\textbf{F1}&\textbf{acc} \\
\hline
\multirow{5}{*}{Our method}&1&	86.83&	90.5&	86.53&	86\\
&5&	94.17&	96.5&	94.27&	93.75\\
&9&	\textbf{94.83}&	96.9&	\textbf{95.85}&	\textbf{94.25}\\
&13&	95&	96.5&	94.7&	94.25\\
&17	&94.33&	97&	94.7&	94.25\\
\hline
\end{tabular}
\label{tab6}
\end{center}
\end{table}

\begin{table}[htbp]
\caption{Influence of ProtoNet out channel dimension on anomaly detection (\%)}
\begin{center}
\begin{tabular}{cccccc}
\hline
\textbf{Method}&\textbf{out dimension}&\textbf{pre}&\textbf{rec}&\textbf{F1}&\textbf{acc} \\
\hline
\multirow{5}{*}{Our method}&16&	94.83&	96.5&	94.63&	94.25\\
&32&	95.17&	\textbf{97.5}&	\textbf{95.57}&	\textbf{95.25}\\
&64&	94.67&	96.5&	94.57&	94.25\\
&128	&93.67&	96.5&	94.07&	93.5\\
&256&	\textbf{96.83}&	94.5&	94.57&	95\\
\hline
\end{tabular}
\label{tab7}
\end{center}
\end{table}
Meanwhile, from Table \ref{tab3}, we can also observe that the performance is sensitive to the parameter $\alpha$, the change of $\alpha$ significantly impact the results. Specifically, adding a regularizer into the prediction loss not only improves the F1 score by 1.77\% but also reduces the FAR by 4.5\%. This demonstrates our robust cost function can really promote the generalization ability of the proposed FSL-PN.

\subsection{Ablation Study}
We also conducted ablation experiments to demonstrate the role of each module of the proposed model in identifying abnormalities on both UNSW\_NB15 and NSL\_KDD datasets. The results are shown in Table \ref{tab4} and Table V, respectively. In both tables, the first record uses only Prototypical Network for anomaly detection. The second and third record shows the experimental results of encoder module+ linear classifier and encoder module + Prototypical Network (ours), respectively. Meanwhile, the CII and SPinfomax denote the proposed class-information injection and MI between sample and corresponding prototype, respectively.

As shown in Table IV and Table V, our model (the third record) gains an increase on F1 score by 2.84\% with respect to the second best option which is encoder module with Liner classifier and by 5.18\% with respect to that using only Prototypical Network for anomaly detection. This can be interpreted as: 1) The proposed feature extractor and the introduced contrastive head are indeed suitable for learning more discriminative feature representations from industrial data; 2) The prototypical network, although can be used alone for few shot learning anomaly detection, its detection performance is limited without an appropriate encoder to help extract the data features. Therefore, combining the prototype network with the proposed encoder not only strengthen the ability of FSL-PN in few-shot learning, but also substantially improves its performance in anomaly detection task with few labeled samples. 

In addition, we also design experiments to verify the effect of convolution layer number of the feature extractor on the model performance, which is shown in Table VI.

The performance of the model increases gradually as the number of convolution layer increases. When the number is nine, the overall performance of the model is optimal. After that, the performance begin to decrease slightly as the number of convolution layer increases. This may be due to the fact that with fewer convolution layers, the dependence between industrial CPS data cannot be adequately modeled while more convolution layers are stacked, the anomalous data may be over-fitted, which leads to the degradation of detection performance. Therefore, the encoder designed in this paper contains only 9 convolution layers in order to adequately learn the high-dimensional features of industrial CPS data while avoiding overfitting on anomalies.

Similarly, we also do experiments to verify the effect of out channel dim of the prototypical network on the detection results, as shown in Table VII. From Table VII, we can see that the model performs best overall when out channel dimension up to 32. This is probably because with a larger out channel dimension, the data features are more scattered and thus less discriminative when dealing with one-dimensional network data streams.

\section{Conclusion}
To guarantee the cyber-physical security of industry applications, this paper proposed FSL-PN to address the issues of imbalanced dataset and few labeled samples produced in CPS. Comparative experiments as well as ablation studies were conducted with existing industry-wide anomaly detection algorithms for few samples and several traditional machine learning methods based on UNSW\_NB15 and NSL\_KDD in few-shot settings. The experimental results show that the proposed method outperforms the existing models in terms of precision, recall, F1 score and FAR metrics, demonstrating the effectiveness of the proposed model in detecting attack signals in industrial CPS environment with few labeled data.

\vspace{12pt}

\end{document}